\documentclass[aps,pre,twocolumn,amsmath,amssymb,showpacs,floatfix]{revtex4}

\usepackage{graphicx}

\renewcommand{\i}{\mathrm{i}}

\newcommand{\T}{\mathrm{T}}
\renewcommand{\L}{\mathrm{L}}
\renewcommand{\k}{\mathbf{k}}

\newcommand{\hopping}{t}
\newcommand{\JH}{J_\mathrm{H}}
\newcommand{\sq}[1]{#1}

\begin{document}

\title{Sequential Tunneling through Molecular Spin Rings}

\author{J\"org Lehmann}
\author{Daniel Loss}
 \affiliation{Department
 of Physics und Astronomy,
 Universit\"at Basel,
 Klingelbergstrasse~82,
 CH-4056~Basel, Switzerland}

\date{\today}

\begin{abstract}
  We consider electrical transport through molecules with
  Heisenberg-coupled spins arranged in a ring structure in the
  presence of an easy-axis anisotropy. The molecules are coupled to
  two metallic leads and a gate.  In the charged state of the ring, a
  Zener double-exchange mechanism links transport properties to the
  underlying spin structure. This leads to a remarkable contact-site
  dependence of the current, which for an antiferromagnetic coupling
  of the spins can lead to a total suppression of the zero-bias
  conductance when the molecule is contacted at adjacent sites.
\end{abstract}


\maketitle

During the last decade, we have witnessed a tremendous progress in the
experimental methods for contacting single molecules and measuring the
electrical current through
them~\cite{Cuniberti2005a,Park1999a,Heersche2006a,Jo2006a,Hirjibehedin2006a}.
A break-through has been achieved with the electromigration-junction
technique~\cite{Park1999a}, which allowed one to add a back-gate to
the molecule, thereby bringing the whole field of Coulomb-blockade
physics to molecular systems.  Recently, this technique has led to the
first transport measurements through single magnetic molecules: The
current through the different Mn$_{12}$ clusters~\cite{Sessoli1993a} has
been measured and magnetic excited states have been
identified~\cite{Heersche2006a, Jo2006a}. At the same time, various
theoretical models for the description of these measurements have been
put forward, all based on the effective Hamiltonians routinely used
for the description of thermodynamic properties of the molecular
magnet~\cite{Heersche2006a, Jo2006a, Romeike2006a, Timm2006a,
  Leuenberger2006a}.  Doing so, an immediate complication encountered
is that for the description of the sequential tunneling processes
observed in the experiment, at least one charged state of the molecule
has to be included in the model, as well.  As plausible models for
this state, the effective Hamiltonian with renormalized
parameters~\cite{Heersche2006a, Romeike2006a} or modified by a single
additional orbital~\cite{Timm2006a} have been employed.

Here, we take a microscopic approach in terms of the constituent spins of the molecular magnet.  Whereas
this limits us to smaller systems than the ones considered
experimentally, it enables us to include an excess charge.
This will shed some light on the role of the
orbital degree of freedom of this excess charge (say an electron)
not contained in an effective-Hamiltonian model.
Further, we are able to consider on the same footing transport
through antiferromagnetically (AF) coupled rings like molecular ferric
wheels~\cite{FerricWheels},
which has not been studied so far.

Evidently, an excess electron will reduce one of the positively
charged ions forming the localized spins. The arbitrary choice of this
ion leads to an orbital degeneracy which will, however, be lifted by a
hopping process between the different sites. At the same time, this
hopping allows the electron to travel from one side of the molecule to
the other and a current to flow across the molecule. Since according
to Hund's rules, the coupling of the excess electron to the localized
spin will favor one direction of the excess electron's spin, and 
assuming that hopping leaves the spin
unchanged~\footnote{For simplicity, we disregard spin-orbit-interaction
induced spin flips during the hopping
  process.}, one obtains an
effective ferromagnetic (FM) type of coupling known as Zener double
exchange~\cite{DoubleExchange}.
Correspondingly, the orbital degree of freedom of the electron is
linked with the spin structure of the ring.
In particular for AF rings this should lead to interesting
transport effects:
Due to the N\'eel-ordered structure of the
AF ground state, we even expect---and indeed confirm---a strong
even-odd-site-dependence of the current mediated by ground-state
transitions not present in the FM case. Another
intriguing property of AF rings is that quantum
tunneling of magnetization~(QTM) leads to a tunnel-split ground-state
doublet with a characteristic oscillating dependence on the magnetic
field~\cite{Chiolero1998a}. We will demonstrate how this behavior can
be extracted in a transport measurement.

\textit{Model}---The physical picture described in the previous paragraph
is captured by the Hamiltonian 
\begin{equation}
  \label{eq:H0}
  \begin{split}
    H_0 = {}&
    -J \sum_{i} \mathbf{S}_i \cdot \mathbf{S}_{i+1}
    -
    \JH
    \sum_{i}
    \sq{\mathbf{s}}_i \cdot \mathbf{S}_i
        - k_z \! \sum_{i} (S^z_i + \sq{s}^z_i)^2
    \\
    &     
    +
    g \mu_\mathrm{B} \, \Big(\mathbf{S} + 
    \sq{\mathbf{s}} 
    \Big) \cdot \mathbf{B}
    - \hopping
    \bigg(
    \sum_{i, \sigma}
    \sq{d}^\dagger_{i,\sigma} \sq{d}^{\vphantom{\dagger}}_{i+1,\sigma}
    + \text{h.c.}
    \bigg)
    \\ & 
    + (\epsilon_0 - e V_\mathrm{g})\, \sq{n}
    + U \sq{n}(\sq{n}-1)/2
    \,.
    \\[-0.45cm]
  \end{split}
\end{equation}
Here, the first term describes the Heisenberg coupling between the
localized spins $\mathbf{S}_i$ with spin quantum number~$s$, where we
require periodic boundary conditions $\mathbf{S}_{N+1} = \mathbf{S}_1$
to describe a ring with total spin $\mathbf{S}=\sum_i \mathbf{S}_i$. The second term represents the Hund's rule
coupling of the excess spin when it is localized at the site $i$ to
the corresponding spin~$\mathbf{S}_i$. The second-quantized spin
operators $\sq{\mathbf{s}}_i$ are defined in terms of the creation
(annihilation) operators $\sq{d}^\dagger_{i,\sigma}$
($\sq{d}_{i,\sigma}$) for an electron of spin $\sigma$ at site $i$ by
$\sq{\mathbf{s}}_{i} = (1/2) \sum_{\sigma\sigma'}
\boldsymbol\tau_{\sigma\sigma'} \, \sq{d}^{\dagger}_{i,\sigma}
\sq{d}_{i,\sigma'} $ with $\boldsymbol\tau$ being the vector of the
three Pauli matrices.  The spin of the excess electron is then given
by $\sq{\mathbf{s}}=\sum_i \sq{\mathbf{s}}_i$.  The easy-axis
anisotropy of strength $k_z$ and the Zeeman energy are included by the
third and fourth term, respectively. The fifth term describes the
hopping of the excess electron around the ring, where again periodic
boundary conditions are assumed.
The orbital energy $\epsilon_0$ of the
excess electron can be tuned by applying an external gate voltage
$V_\mathrm{g}$ and is proportional to the occupation number
$\sq{n}=\sum_{i} \sq{d}^\dagger_{i,\sigma} \sq{d}_{i,\sigma}$. Finally, we restrict
ourselves to a single excess electron by letting the charging energy
$U\to\infty$.

The coupling of the molecule to the metallic leads is described by the
tunnel Hamiltonian $ H_\T = \sum_{\ell \mathbf{k} \sigma} T_{\ell
  \mathbf{k}} \, \sq{c}^\dagger_{\ell \mathbf{k}\sigma} \sq{d}_{i_\ell
  \sigma} + \text{h.c.}$ Here, the sum runs over the two
leads~$\ell=\mathrm{L}, \mathrm{R}$, which we assumed to be each
coupled via a spin-independent coupling matrix elements
$T_{\ell\mathbf{k}}$ to a single ring site~$i_\ell$.  The leads at an
electro-chemical potential $\mu_\ell$ are modeled as a set of
independent quasiparticles with wave-vector $\mathbf{k}$ and spin
$\sigma$, $H_\L = \sum_{\ell\k\sigma} \epsilon_{\ell \mathbf{k}} \,
\sq{c}^\dagger_{\ell\mathbf{k}\sigma} \sq{c}_{\ell\mathbf{k}\sigma}$,
distributed according to the Fermi distribution $f(\epsilon-\mu_\ell)
= \{1+\exp[(\epsilon-\mu_\ell)/kT]\}^{-1}$. Assuming that the
externally applied bias voltage $V_\mathrm{b}$ drops completely and
symmetrically along both molecule-lead contacts, we have
$\mu_{\mathrm{L},\mathrm{R}} = \mu \pm e V_\mathrm{b}/2$, where $\mu$
is the Fermi energy.

\textit{Magnetic order}---The eigenstates $|\alpha\rangle$ of the
unperturbed ring Hamiltonian~\eqref{eq:H0} can be separately
considered for the cases of an uncharged ($n=0$) and charged ($n=1$)
ring.  Moreover, the discrete translational symmetry and the symmetry
of reflection at an arbitrary site, say $i=1$, yield corresponding
good quantum numbers $\cos(q)$ with $-\pi<q\le\pi$ and $N q = 0
\mathop\mathrm{mod} 2\pi $, and parity $P=\pm1$, respectively. In the
absence of a magnetic field with $x$- and $y$-components, another good
quantum number $M$ is given by the $z$-component of the total angular
momentum~$\mathbf{S} + \mathbf{s}$.  The
transition matrix elements
$\langle\alpha'|d^\dagger_{i,\sigma}|\alpha\rangle$ of the creation
operator fulfil the selection rules $n'-n=1$ and $M'-M = \sigma$.
In the special case $N=4$ considered below, we find
$P'=P$ for odd site numbers $i=1,3$ and
$\exp(2\mathrm{i}q')P'=\exp(2\mathrm{i}q)P$ for even site numbers
$i=2,4$.

\begin{figure}[tb]
  \vspace{0.5cm}

  \centering
  \raisebox{5.0cm}{(a)} \includegraphics[width=6.1cm]{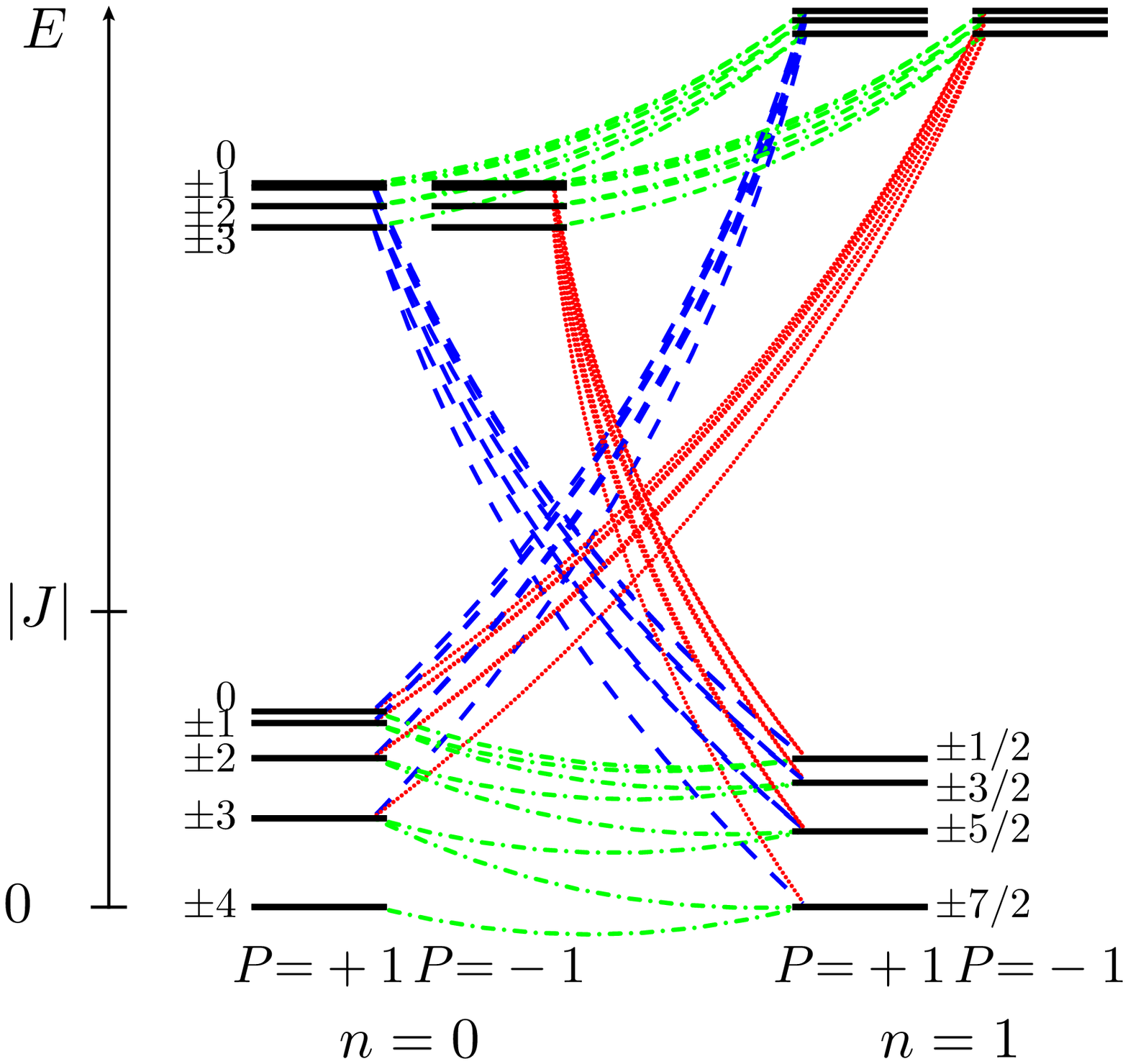}
    
  \vspace{0.1cm}

  \raisebox{5.1cm}{(b)} \includegraphics[width=6.1cm]{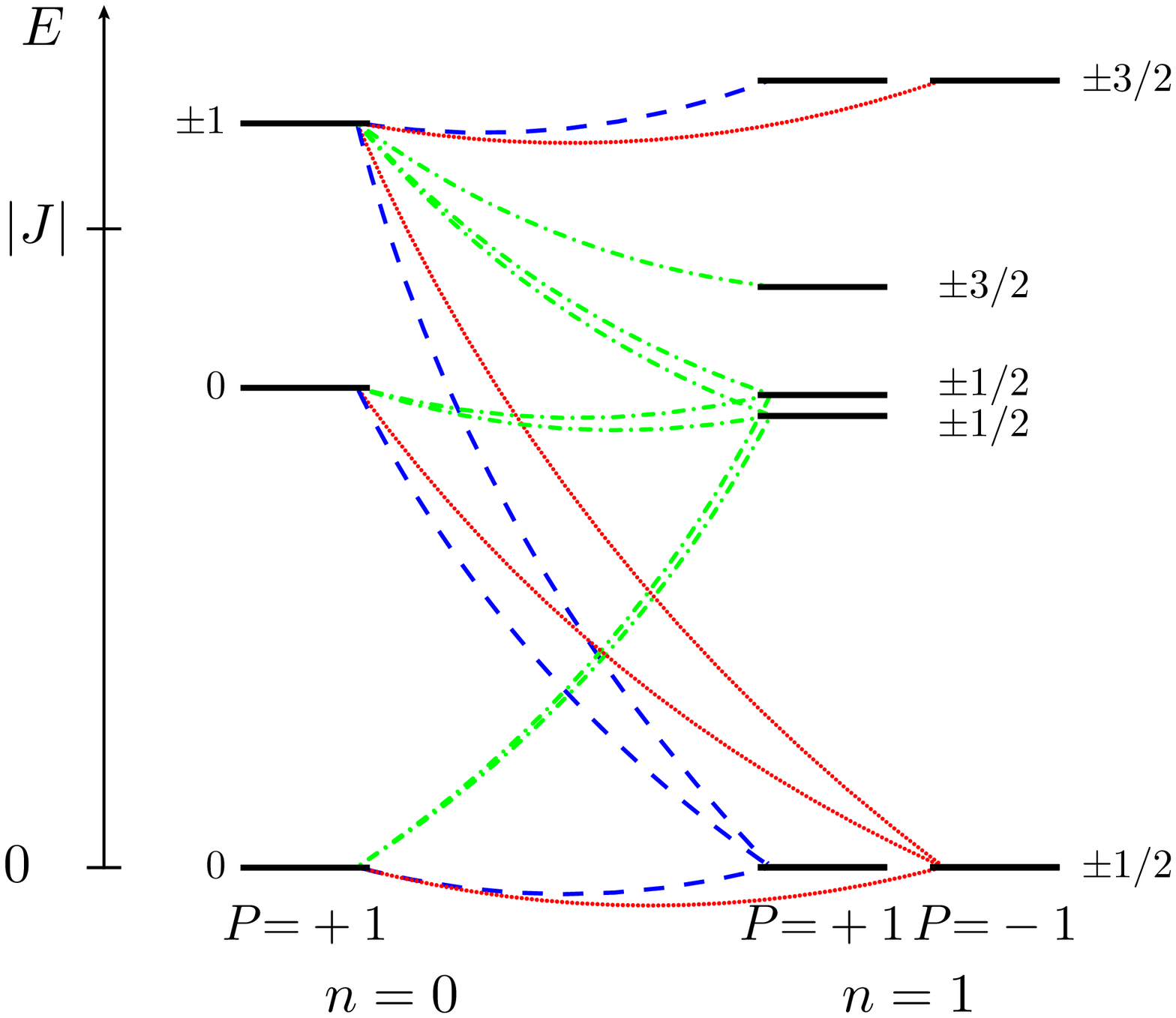}
  \caption{(Color online) Low-energy states of the neutral (left, $n=0$) and charged
    (right, $n=1$) ring both for (a) FM and (b)
    AF coupling. The parameters are given in the main text. The
    levels are labeled by their magnetic quantum number $M$ and their
    parity $P$.  Dashed (dotted) lines indicate tunnel transitions
    allowed when the ring is contacted at the odd (even)
    numbered sites. Transitions allowed for all contact
    sites are drawn as dash-dotted lines.}
  \label{fig:levels}
\end{figure}

We now discuss the magnetic order of the lowest-lying states for the
two opposite cases of a FM and AF intra-ring
coupling~$J$.  In both cases, we assume that the ions forming the
localized spins have at least half-filled shells and thus consider
according to Hund's rules a strong AF coupling
$\JH\ll-|J|$. This describes, for instance, ferric wheels containing
Fe$^{3+}$ ions. We will exemplify the general discussion by the
special case of a ring consisting of $N=4$ spins $s=1$ (cf.\
Fig.~\ref{fig:levels}).
The exchange coupling strength~$|J|$ serves as energy unit
(typical values for ferric wheels range from $20$ to
$30\,k_\mathrm{B}\mathrm{K}$ \cite{FerricWheels}).  The anisotropy is
set to $k_z = 0.3 |J|$ and the Hund's rule coupling strength is
$\JH=-100|J|$.  An estimate for the hopping matrix element~$\hopping$ is
difficult to make. Assuming that it is of the same
order of magnitude as the one leading to exchange in the uncharged ring,
a simple argument based on a Hubbard model yields the lower bound $J/4t
\approx \hopping/U' \ll 1$, where $U'$ is the on-site charging energy.
Expecting the excess electron to be rather localized, we assume
$\hopping=5|J|$.

For a FM intra-ring coupling, the lowest-lying states of
both the neutral and charged ring consist of multiplets of total spin
$S=Ns$ and $S=Ns-1/2$, respectively [cf.\ Fig.~\ref{fig:levels}(a)].
These multiplets are split by the easy-axis anisotropy into doublets with
same absolute value of the magnetic quantum number~$|M|$. The
additional FM-type interaction due to the Zener double
exchange influences the magnetic order only weakly.

This is in stark contrast to the case of an AF
intra-ring coupling~$J<0$ [cf.\ Fig.~\ref{fig:levels}(b)], which, for
the charged ring, competes with the double-exchange contribution. In
order to understand the resulting magnetic ordering in this case, we
first consider the \textit{uncharged} ring. The classical ordering is
described by the two N\'eel states $|{\Uparrow}\rangle = |s,-s,
s,-s,\dots\rangle$ and $|{\Downarrow}\rangle = |-s,s,
-s,s,\dots\rangle$, where the vectors have been written in the
$\{S_{z,i}\}$ eigenbasis.  In the QTM regime defined by the scaled
tunnel action $\mathcal{S}_0/\hbar=N s \sqrt{2k_z/J}$ being much
larger than unity~\cite{Chiolero1998a}, the leading contribution to
the two lowest-lying quantum-mechanical eigenstates of the uncharged
ring is then given by the symmetric and antisymmetric superposition of
the two N\'eel states, i.e., $(|{\Uparrow}\rangle
\pm|{\Downarrow}\rangle)/\sqrt{2}$.

Similarly, in a purely classical picture and without hopping
contribution, the \textit{charged} ring has a $2N$-fold degenerate
ground state which for an AF on-site coupling $\JH<0$
consists of the states $|{\Uparrow}\rangle |i,{\uparrow}\rangle$ and
$|{\Downarrow}\rangle |i,{\downarrow}\rangle$ for odd~$i$ as well as
$|{\Uparrow}\rangle |i,{\downarrow}\rangle$ and $|{\Downarrow}\rangle
|i,{\uparrow}\rangle$ for even~$i$.  Here, $|i,\sigma\rangle$
describes the state of the excess electron being localized at site~$i$
with spin $\sigma$. The hopping leads to an hybridization of the
states of same N\'eel ordering and excess electron spin.
Furthermore, QTM couples states with opposite N\'eel ordering of the
localized spin lattice.

We find for the special case with $N=4$ and $s=1$ [cf.\
Fig.~\ref{fig:levels}(b)] that the ground and first excited state of
the uncharged ring are indeed dominated by the superposition of the
N\'eel-ordered states described above. The ground state of the charged
ring turns out to be four-fold degenerate, discerning into two
degenerate Kramers doublets with $M=\pm1/2$.  Focusing on the states
with $M=1/2$, we find that in the parity ($P$) eigenbasis, the main
contribution to the eigenstates stems from the states
$|{\Uparrow}\rangle (|2,\uparrow\rangle - |4,\uparrow\rangle)$ with
positive parity~$P=1$ and $|{\Downarrow}\rangle (|1,\uparrow\rangle -
|3,\uparrow\rangle)$ with negative parity~$P=-1$. These states do not
exhibit a QTM-induced tunnel splitting~\footnote{We speculate that a
  physical interpretation might be given in terms of a
  geometrical-phase
  effect~\cite{BerryPhase}. However,
  this needs further investigation.}. The situation is different for
the two next excited Kramers doublets. The leading contribution to
these states is given by the symmetric and antisymmetric combination
of the two states $|{\Uparrow}\rangle
(|1,\uparrow\rangle+|3,\uparrow\rangle)$ and $|{\Downarrow}\rangle
(|2,\uparrow\rangle+|4,\uparrow\rangle)$ and we indeed observe a
finite tunnel-splitting due to QTM.

\textit{Bloch-Redfield approach}---We calculate the sequential-tunneling current through
the ring using a Bloch-Redfield approach, extending previous results~\cite{BlochRedfield}.
In contrast to a Pauli
master-equation description, this includes off-diagonal elements
(coherences) of
the reduced density-matrix $\varrho_{\alpha\beta} = \mathrm{Tr}_\mathrm{L} \langle
\alpha|\varrho|\beta\rangle$ after tracing out the leads and is 
also valid for level spacing less than the level broadening induced by
the leads. We find
\begin{equation}
  \label{eq:master}
  \begin{split}
  \dot{\varrho}_{\alpha\beta} = &
  -\i \omega_{\alpha\beta} \varrho_{\alpha\beta}
  + \frac{1}{2}
  \sum_{\ell \alpha' \beta'} \!
  \bigg\{\!\!
  \left[W^\ell_{\beta' \beta \alpha \alpha'}  + (W^\ell_{\alpha' \alpha \beta \beta'})^\ast\right]
  \varrho_{\alpha'\beta'}\\
  &\quad-
  W^\ell_{\alpha \beta'\beta'\alpha'}\,
  \varrho_{\alpha'\beta}
  -
  (W^\ell_{\beta \alpha'\alpha'\beta'})^\ast\,
  \varrho_{\alpha\beta'}
  \bigg\}\,.
  \\[-0.65cm]
  \end{split}
\end{equation}
Here, we have introduced the frequencies
$\omega_{\alpha\beta}=(E_\alpha-E_\beta)/\hbar$ and the transition rates
$W^\ell_{\beta' \beta \alpha \alpha'} = W^{\ell+}_{\beta' \beta \alpha
  \alpha'} + W^{\ell-}_{\beta' \beta \alpha \alpha'}$ due to tunneling
across a molecule-lead contact $\ell$ in terms of the rates 
$W^{\ell+}_{\beta' \beta \alpha \alpha'} = \Gamma_{\ell}(E_\alpha-E_{\alpha'}) \,
f(E_\alpha-E_{\alpha'}-\mu_\ell) \sum_{\sigma} \langle
\beta'|d_{i_\ell,\sigma}|\beta\rangle\!\langle
\alpha|d_{i_\ell,\sigma}^\dagger|\alpha'\rangle$ and $W^{\ell-}_{\beta'
  \beta \alpha \alpha'} = \Gamma_{\ell}(E_{\alpha'}-E_{\alpha}) \,
[1-f(E_{\alpha'}-E_{\alpha}-\mu_\ell)] \sum_{\sigma}  \langle
\beta'|d^\dagger_{i_\ell\sigma}|\beta\rangle\!\langle
\alpha|d_{i_\ell,\sigma}|\alpha'\rangle$ for
tunneling of an electron on and off the molecule, where
$\Gamma_{\ell}(\epsilon) = (2\pi/\hbar) \sum_{\mathbf{k}}
|T_{\ell \mathbf{k}}|^2 \delta(\epsilon-\epsilon_{\ell\mathbf{k}})$ is
the spectral density of the coupling. Finally, the current
across contact $\ell$ is given by
$ I_\ell = e\, \mathrm{Re} \sum_{\alpha \alpha' \beta}
\big( W^{\ell-}_{\beta \alpha' \alpha' \alpha} - W^{\ell+}_{\ell \beta
  \alpha' \alpha' \alpha} \big)\varrho_{\alpha\beta}$.
 
The number of rates~$W^\ell_{\beta' \beta \alpha \alpha'}$ scales with the fourth
power of the dimension of the molecular Hilbert space. For the numerical
solution of Eq.~\eqref{eq:master}, we thus cannot take
into account all coherences. However, for the stationary solution, coherences between states with an energy
difference much larger than the molecule-lead coupling strength are
negligible.

\begin{figure}[tb]
  \centering
    \includegraphics[width=0.9\linewidth]{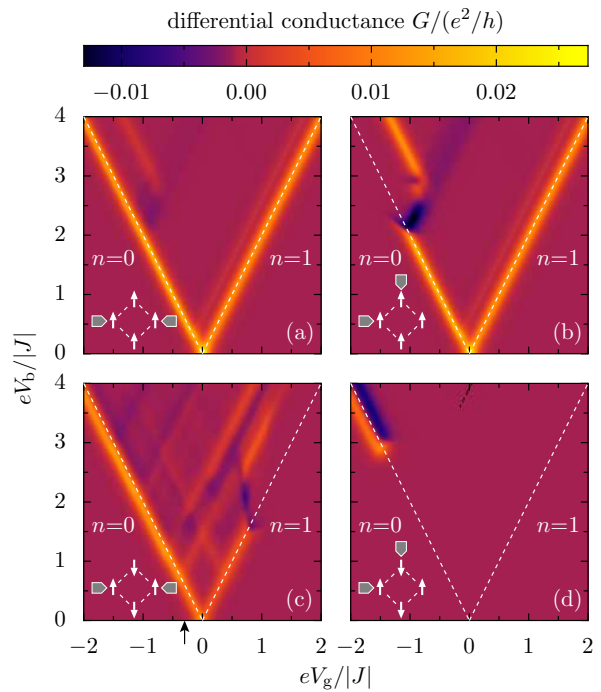}
    \caption{(Color online) Differential conductance (color-coded) as a function of
      gate and bias voltage at zero magnetic field. The temperature is
      $k_\mathrm{B} T = 0.05 |J|$ and the other parameters are as
      given in the main text. The upper two panels (a) and (b) show
      the conductance for an FM coupling $J>0$, the lower
      panels (c) and (d) depict the AF situation $J<0$.
      The left (right) panels show the conductance when the molecule
      is contacted at opposite (adjacent) sites $i_\mathrm{L}=1$ and
      $i_\mathrm{R}=2$ ($i_\mathrm{L}=1$ and $i_\mathrm{R}=3$) (cf.
      inset schematic drawings). The dashed lines indicate the
      position of the ground-state transitions. }
  \label{fig:af-f}
\end{figure}

\textit{Differential conductance}---For our numerical calculations, we
use the model parameters described above and assume a
contact- and energy-independent molecule-lead coupling~$\hbar\Gamma=\hbar\Gamma_{\ell}=0.01|J|$.
Figure~\ref{fig:af-f} shows the differential conductance~$G=dI/dV_\mathrm{b}$ as a function
of bias and gate voltage at zero magnetic field for both the FM
and the AF case and for two molecule-lead
coupling variants: contacts at \textit{opposite} and \textit{adjacent}, respectively,
sides of the ring.
For a FM coupling~$J>0$, the behavior in the low-bias-voltage
regime is identical in both cases [see Figs.~\ref{fig:af-f}(a) and
(b)]: the ground-state transitions dominate the current and one
excited state can be identified.
Away from the charge-degeneracy point, one ground-state transition
is suppressed for adjacent contacts, for reasons similar to the
AF case discussed next~\cite{we}.

\begin{figure}[tb]
  \centering
  
  \vspace{0.5cm}

  \includegraphics[width=0.60\linewidth]{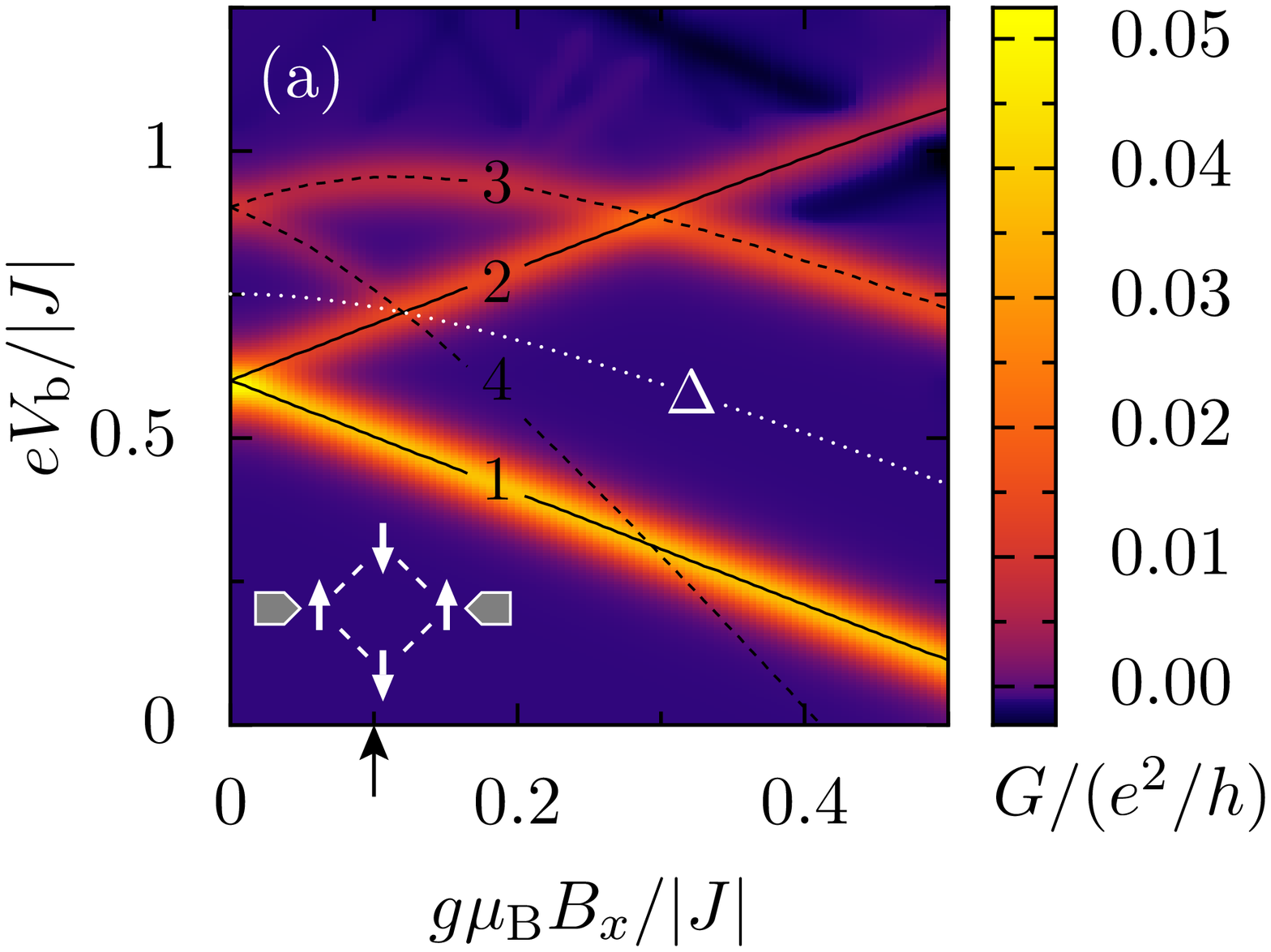}
  \raisebox{0.3cm}{\includegraphics[width=0.38\linewidth]{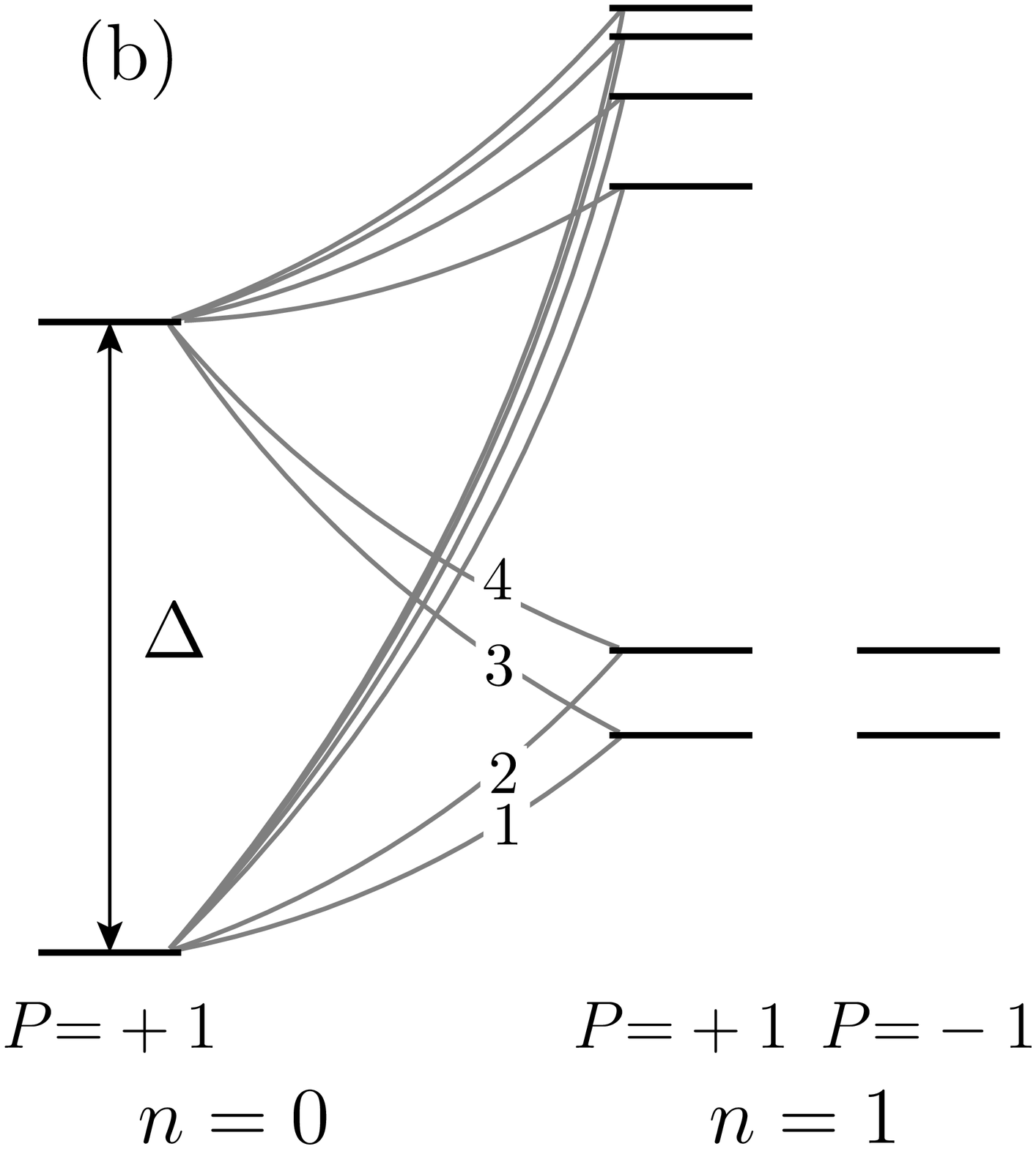}}
  \caption{(Color online) (a) Differential conductance through an
    AF coupled ring contacted at opposite sides as
    a function of magnetic field in $x$-direction and bias voltage.
    The gate voltage has been chosen as indicated by the arrow in
    Fig.~\ref{fig:af-f}(d). The temperature is $k_\mathrm{B}T = 0.01
    |J|$ and the other parameters are given in the main text. The
    solid (dashed) lines indicate the electron (hole) transition
    labeled in panel~(b). The tunnel splitting~$\Delta$ extracted from
    these transitions is shown as white dotted line. (b) Level
    structure of the ring and tunnel transitions for finite $B_x$ as
    indicated by the arrow in panel~(a).}
  \label{fig:bx}
\end{figure}

In the case of an AF intra-ring coupling~$J<0$, we find
a more striking difference between the two contacting situations.  For
adjacent contacts [see Fig.~\ref{fig:af-f}(d)], \textit{the low-bias
conductance is completely suppressed}, which does not occur for
contacts at opposite sides [see Fig.~\ref{fig:af-f}(c)].  In order to
understand this remarkable difference, we consider the magnetic order
discussed above. There, we identified two Kramers doublets of different
parity in the $n=1$ ground state. Furthermore we found the contact-site
dependent selection
rules depicted in Fig.~\ref{fig:levels}. In particular, when contacting at adjacent
sites, one contact only allows transitions to the even parity states
and the other one only to the ones with odd parity. Thus, there is no
transport path connecting both contacts and the current is blocked
completely until transitions via excited states become possible at
higher bias-voltages.
On the other hand, for contacts at opposite sides, only
one parity state becomes populated and a current can flow. It displays a rather complex excited-state spectrum compared to
the FM case. For a better understanding of this regime, we
focus on a fixed gate voltage as indicated by the arrow in
Fig.~\ref{fig:af-f}(c) and consider in Fig.~\ref{fig:bx}(a) the
conductance as a function of an in-plane magnetic field.
This dependence is
particularly interesting for AF rings where the QTM-induced tunnel
splitting $\Delta$ between ground and first
excited state of the neutral ring [see Fig.~\ref{fig:bx}(b)] oscillates
as a function of the magnetic field~\cite{Chiolero1998a}.
In order to
extract this splitting from the conductance behavior, we identify all
transitions between the lowest-lying states of the neutral and charged
molecule (see labeled lines in Fig.~\ref{fig:bx}). Denoting the
magnetic-field dependent bias voltages at the corresponding resonances
by $V_{\mathrm{b},k}$ ($k=1,\dots,4$), the tunnel splitting can be
obtained as $\Delta = e(V_{\mathrm{b},1}+V_{\mathrm{b},3})/2 =
e(V_{\mathrm{b},2}+V_{\mathrm{b},4})/2$.

We finally remark that a similar contact dependence of the current, albeit not
a complete suppression of the zero-bias conductance, has been
predicted for benzene rings~\cite{Hettler2003a}. There, it was
also noted that when excited states are populated, a non-tunneling
decay may lead to a population of the odd-parity state such that for
contacts at opposite sides one may find a current collapse at higher
bias voltages. A similar scenario can be envisaged here with the
additional possibility to lift the blockade again by applying a
large-enough magnetic field~$B_x$~\cite{we}. Then, the current should
display an interesting hysteric behavior possibly useful 
for information storage purposes.

To conclude, we have presented a model for a molecular
spin ring in the presence of a single excess electron. It
contains a hopping term which both provides a transport path and
results in a Zener double-exchange coupling between the localized
spins.  For AF coupled rings of ions with least
half-filled shells, the interplay of these two effects leads to a
strong contact-dependence of the sequential tunneling current through
the molecule reflecting the N\'eel-ordered ground states.

We thank D.~Bulaev, W.A.~Coish, J.~Egues, and M.~Wegewijs
for discussions. Financial support by the EU RTN QuEMolNa,
the NoE MAGMANet, the NCCR Nanoscience, and the Swiss
NSF is acknowledged.


\begin{thebibliography}{10}

\bibitem{Cuniberti2005a}
{\em Molecular Electronics}, {\em Lecture Notes in Physics}, edited by G.
  Cuniberti, G. Fagas, and K. Richter (Springer, Berlin, 2005).

\bibitem{Park1999a}
H. Park {\it et~al.}, Appl. Phys. Lett. {\bf 75},  301  (1999).

\bibitem{Heersche2006a}
H.~B. Heersche {\it et~al.}, Phys. Rev. Lett. {\bf 96},  206801  (2006).

\bibitem{Jo2006a}
M.-H. Jo {\it et~al.}, Nano Lett. {\bf 6},  2014  (2006).

\bibitem{Hirjibehedin2006a}
C.~F. Hirjibehedin, C.~P. Lutz, and A.~J. Heinrich, Science {\bf 312},  1021
  (2006).

\bibitem{Sessoli1993a}
R. Sessoli {\it et~al.}, Nature (London) {\bf 365},  141  (1993).

\bibitem{Romeike2006a}
C. Romeike, M.~R. Wegewijs, and H. Schoeller, Phys. Rev. Lett. {\bf 96},
  196805  (2006).

\bibitem{Timm2006a}
C. Timm and F. Elste, Phys. Rev. B {\bf 73},  235304  (2006).

\bibitem{Leuenberger2006a}
M.~N. Leuenberger and E.~R. Mucciolo, Phys. Rev. Lett. {\bf 97},  126601
  (2006).

\bibitem{FerricWheels}
K.~L. Taft {\it et~al.}, J. Am. Chem. Soc. {\bf 116},  823  (1994);
A. Caneschi {\it et~al.}, Chem. Eur. J. {\bf 2},  1379  (1996);
M. Affronte {\it et~al.}, Phys. Rev. B {\bf 60},  1161  (1999);
B. Normand {\it et~al.}, \textit{ibid.} {\bf 63},  184409  (2001);
A. Honecker {\it et~al.}, Eur. Phys. J. B {\bf 27},  487  (2002).

\bibitem{DoubleExchange}
C. Zener, Phys. Rev. {\bf 82},  403  (1951);
P.~W. Anderson and H. Hasegawa, \textit{ibid.} {\bf 100},  675  (1955).

\bibitem{Chiolero1998a}
A. Chiolero and D. Loss, Phys. Rev. Lett. {\bf 80},  169  (1998).

\bibitem{BlochRedfield}
H.-A. Engel and D. Loss, Phys. Rev. Lett. {\bf 86},  4648  (2001);
S. Kohler, J. Lehmann, and P. H\"anggi, Phys. Rep. {\bf 406},  379  (2005).

\bibitem{we}
Details will be presented elsewhere.

\bibitem{Hettler2003a}
M.~H. Hettler {\it et~al.}, Phys. Rev. Lett. {\bf 90},  076805  (2003).

\bibitem{BerryPhase}
D. Loss, D.~P. DiVincenzo, and G. Grinstein, Phys. Rev. Lett. {\bf 69},  3232
  (1992);
J. von Delft and C.~L. Henley, \textit{ibid.} {\bf 69},  3236  (1992);
M.~N. Leuenberger and D. Loss, Phys. Rev. B {\bf 63},  054414  (2001).

\end{thebibliography}
\end{document}